\documentclass[aps,amsmath,showpacs,amsfonts,10pt]{revtex4}
\usepackage{epsfig,graphicx}
\usepackage[english]{babel}
\usepackage{amsfonts}
\usepackage{amsmath}
\usepackage{latexsym}
\usepackage{graphics,bm}
\usepackage{natbib}
\usepackage{dcolumn}
\usepackage{bm}
\usepackage{rotating}
\begin{document}

\title{Selective entanglement in a two-mode optomechanical system}

\author{Neha Aggarwal$^{1,2}$, Kamanasish Debnath$^{1}$, Sonam Mahajan$^{1}$, Aranya B. Bhattacherjee$^{2,3}$ and Man Mohan$^{1}$}

\address{$^{1}$Department of Physics and Astrophysics, University of Delhi, Delhi-110007, India} \address{$^{2}$Department of Physics, ARSD College, University of Delhi (South Campus), New Delhi-110021, India}\address{$^{3}$School of Physical Sciences, Jawaharlal Nehru University, New Delhi-110067, India}

\begin{abstract}

We analyze an optomechanical system formed by a mechanical mode and the two optical modes of an optomechanical cavity for the realization of a strongly quantum correlated three-mode system. We show that the steady state of the system shows three possible bipartite continuous variable entanglements in an experimentally accessible parameter regime, which are robust against temperature. We further show that selective entanglement between the mechanical mode and any of the two optical modes is also possible by the proper choice of the system parameters. Such a two-mode optomechanical system can be used for the realization of continuous variable quantum information interfaces and networks.

\end{abstract}


\maketitle

\section{Introduction}

The field of optomechanics of micro- and nanocavities has recently sparkled the interest of a vast scientific community due to its distinct applications, ranging from sensing of masses, forces, and displacements at the ultimate quantum limits \citep{sch,kip}, to tests of the validity of quantum mechanics at a macroscopic level \citep{mar,rom}, up to the realization of quantum interfaces for quantum information networks \citep{man1,pir,hammerer,rab}. Braginski and his co-workers \citep{bra} first pointed out the possibility to detect genuine quantum behavior in cavities characterized by radiation pressure interaction between an optical mode and a mechanical resonator in the direction of the interferometric detection of gravitational waves. Although, earlier, several different schemes have been established to detect quantum mechanical effects in such systems, such as continuous variable (CV) entanglement between cavity modes and/or mechanical modes, squeezed states of the light or the mechanical modes, and ground-state cooling of the mechanical modes \citep{gen}. Such schemes include cavities and resonators at the micro- or nano-level rather than the macroscopic scale of gravitational wave detectors. The unique opportunities to engineer optomechanical devices is due to the profit obtained from the tremendous progress in micro- and nano-fabrication techniques. A variety of examples are toroidal optical microresonators \citep{kip}, Fabry-Perot cavities with a movable micro-mirror \citep{gigan,arcizet}, a semi-transparent membrane within a standard Fabry-Perot cavity \citep{tho,jay,wil,san}, suspended silicon photonic waveguides \citep{li,li1,li2}, SiN nanowires evanescently coupled to a microtoroidal resonator \citep{anetsberger}, adjacent photonic crystal wires \citep{eic}, nano-electromechanical systems formed by a microwave cavity capacitively coupled to a nano-resonator \citep{teu,roc,teu1}, and atomic ensembles interacting with the optical mode of an optical cavity \citep{bre,mur}.

Entanglement is one of the characteristic elements of quantum theory as it is responsible for correlations between observables that cannot be understood on the basis of local realistic theories \citep{bel}. It is now intensively studied in performing communication and computation tasks with an efficiency that cannot be achieved classically \citep{nielsen}. This leads to an outpouring interest in establishing the conditions under which entanglement between macroscopic objects is possible. A useful and comprehensive review of the theory of entanglement in systems of continuous variables was concisely given by Eisert and Plenio in \citep{eisert1}. Relevant experimental demonstrations in this direction are given by the entanglement between collective spins of atomic ensembles \citep{jul}, and between Josephson-junction qubits \citep{ber}. Further, two mirrors of a ring cavity are entangled by the radiation pressure of the cavity mode \citep{man}. Many other proposals involve nano- and micro-mechanical resonators entangled with other systems. A nanomechanical oscillator could be entangled with a Cooper-pair box \citep{arm}. However, Ref. \citep{eis} studied how to entangle an array of nanomechanical oscillators. In addition to this, many more proposals suggest the entanglement between two charge qubits \citep{zou} or two Josephson junctions \citep{cle} via nanomechanical resonators, or to entangle two nanomechanical resonators via trapped ions \citep{tia}, Cooper pair boxes \citep{tia1}, or dc-SQUIDS \citep{xue}. Furthermore, a system of two mirrors of two different cavities illuminated with entangled light beams is studied \citep{zha}. Moreover, Refs. \citep{pin,pat,bha,wip} demonstrated different examples of double-cavity systems in which entanglement is examined either between different mechanical modes, or between a cavity mode and a vibrational mode of a cavity mirror. 

The simplest scheme capable of generating stationary optomechanical entanglement is studied with a single movable Fabry-Perot cavity mirror \citep{vitali1}, or both movable cavity mirrors \citep{vitali}. The generation of stationary entanglement from the Fabry-Perot model of Ref. \citep{vitali1} is remarkable for its simplicity and robustness against temperature. In fact, entangled optomechanical systems have potential profitable application in realizing quantum communication networks, in which the mechanical modes play the vital role of local nodes where quantum information can be stored and retrieved, and optical modes carry this information between the nodes. This type of scheme is proposed in Refs. \citep{man1,pir,pir1,pir2} which is based on free-space light modes scattered by a single reflecting mirror. This allows the implementation of continuous variable (CV) quantum teleportation \citep{man1,pir2}, quantum telecloning \citep{pir}, and entanglement swapping \citep{pir1}. Further, entanglement in the steady state of a system could be significantly useful for quantum communication applications since it is stationary, i.e., it has a virtually infinite lifetime, and hence could be used repeatedly. Due to this reason, it allows more robust uses of entanglement, at variance with the plethora of schemes where entanglement is obtained only after a given interaction time and has a finite lifetime \citep{yu}.

Motivated by these interesting developments in the field of optomechanics, we consider an optomechanical system consisting of two optical modes coupled to a mechanical resonator through the radiation pressure interaction. The two optical modes are also coupled together via the common interaction with the mechanical oscillator. It has been demonstrated experimentally that the quantum nonlinearity can be enhanced significantly by coupling two optical modes to a mechanical resonator \citep{safavi,ludwig,thompson,grudinin}. This enhanced nonlinearity has potential applications in quantum nondemolition (QND) measurements \citep{grangier} and quantum information processing \citep{stannigel}. The selective energy exchange between any two modes of such an optomechanical system has also been investigated very recently \citep{neha}. Here, we study this system to show how the stationary bipartite CV entanglements can be generated between the different modes of the system. These steady-state entanglements are quantified by the logarithmic negativity and are further shown to be robust against temperature. We also demonstrate the possibility of selective entanglement between the mechanical mode and any of the two optical cavity modes of the system by choosing an appropriate parameter regime. The variation in optomechanical entanglement with the change in its respective optomechanical coupling is also studied. Such entangled optomechanical systems can be profitably used for the realization of CV quantum-communication networks.

\section{The Model}

In this section, we introduce the basic model for our system whose schematic representation is depicted in fig.(1). The optomechanical system considered here consists of a single mechanical mode (denoted by operator $b$) which couples the two optical modes (denoted by operators $c_{1}$ and $c_{2}$) of an optomechanical cavity via radiation pressure. The simplest description of the radiation-pressure coupling of the mechanical and optical modes in such a system can be provided by the following Hamiltonian \citep{neha}:

\begin{equation}\label{ham}
H=\hbar \omega_{c_{1}}c_{1}^{\dagger}c_{1}+\hbar \omega_{c_{2}}c_{2}^{\dagger}c_{2}+\hbar \omega_{m}b^{\dagger}b-\hbar \eta_{1}c_{1}^{\dagger}c_{1}(b+b^{\dagger})-\hbar \eta_{2}c_{2}^{\dagger}c_{2}(b+b^{\dagger})+\hbar \eta_{0}(c_{1}c_{2}^{\dagger}+c_{2}c_{1}^{\dagger})(b+b^{\dagger}).
\end{equation}

Here, the first and second terms in the Hamiltonian represent the bare energies of the two optical modes. For each of the two optical modes of the cavity, we associate an annihilation (creation) operator $c_{r}(c_{r}^{\dagger})$ $([c_{r},c_{r}^{\dagger}]=1)$ and a frequency $\omega_{c_{r}}(r=1,2)$. Third term gives the energy of the single vibrational mode of the mechanical resonator having frequency $\omega_{m}$ with a lowering (raising) operator $b(b^{\dagger})$ such that $[b,b^{\dagger}]=1$. Fourth term illustrates the interaction between the optical mode ($c_{1}$)and the mechanical mode with coupling rate $\eta_{1}$. Fifth term in the Hamitonian depicts the coupling between the cavity mode ($c_{2}$) and the mechanical resonator, which is denoted by $\eta_{2}$. Such type of interaction results from the space dependence of the cavity frequencies. Last term in the Hamiltonian gives the energy due to coupling between the position quadrature of the mechanical oscillator and the two optical cavity modes. Such an optomechanical system involving two optical modes coupled to a mechanical mode is analogous to a travelling-wave phonon-photon translator (PPT) \citep{safavi}. In such devices, one of the cavity modes enables the inter-conversion of phonons in the mechanical mode to photons in the second optical cavity mode via two-photon process. The optomechanical coupling rate between the two optical modes via mechanical resonator is denoted by $\eta_{0}$, which basically corresponds to the coherent Kerr-type interaction responsible for introducing the quantum nonlinearities into the system \citep{ludwig}. Here, we assumed the mechanical displacements to be small enough such that the linear order becomes the only important term in the interaction. A Hamiltonian of this kind is found typically in the "membrane in the middle"-setup \citep{thompson}.  

\begin{figure}[h]
\hspace{-0.0cm}
\includegraphics [scale=0.8]{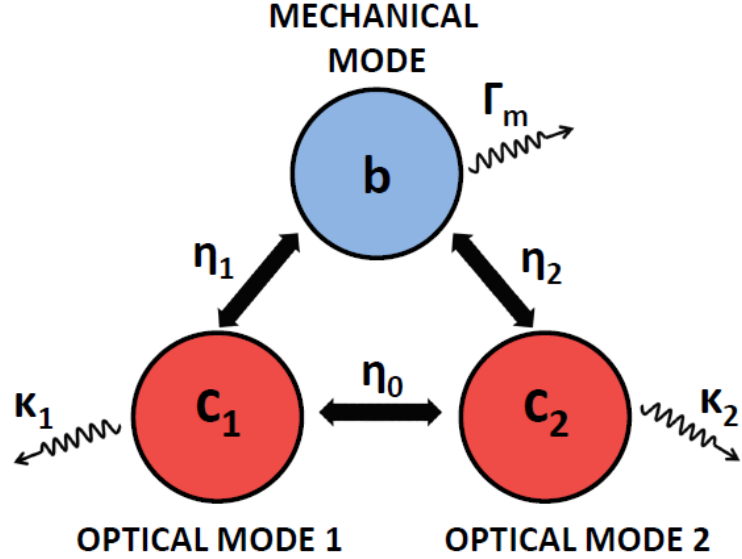}
\caption{(color online) Schematic representation of the optomechanical system consisting of two optical modes $c_{1}$ and $c_{2}$ and one mechanical mode $b$. Also shown are the coupling rates ($\eta_{0}$, $\eta_{1}$ and $\eta_{2}$) between the various modes with the different damping rates ($\kappa_{1}$, $\kappa_{2}$ and $\Gamma_{m}$) corresponding to each of the bosonic modes.}
\end{figure}\label{fig1}

Using the Hamiltonian given by eqn.(\ref{ham}), the dynamics of the system can be described by the following set of coupled nonlinear quantum Langevin equations (QLEs) for the bosonic field operators $b(t)$, $c_{1}(t)$ and $c_{2}(t)$:

\begin{equation}\label{b}
\dot{b}(t)=-i\omega_{m}b(t)+i\eta_{1}c_{1}^{\dagger}(t)c_{1}(t)+i\eta_{2}c_{2}^{\dagger}(t)c_{2}(t)-i \eta_{0} (c_{1}(t)c_{2}^{\dagger}(t)+c_{2}(t)c_{1}^{\dagger}(t))-\Gamma_{m}b(t)+\sqrt{\Gamma_{m}}\xi_{m}(t),
\end{equation}

\begin{equation}\label{c1}
\dot{c_{1}}(t)= -i \omega_{c_{1}} c_{1}(t)+ i \eta_{1} c_{1}(t) (b(t)+b^{\dagger}(t))-i \eta_{0} c_{2}(t) (b(t)+b^{\dagger}(t)) -\frac{\kappa_{1}}{2}c_{1}(t)+ \sqrt{\kappa_{1}} c_{in_{1}}(t),
\end{equation}

\begin{equation}\label{c2}
\dot{c_{2}}(t)=-i\omega_{c_{2}}c_{2}(t)+i\eta_{2}c_{2}(t)(b(t)+b^{\dagger}(t))-i\eta_{0}c_{1}(t)(b(t)+b^{\dagger}(t))-\frac{\kappa_{2}}{2}c_{2}(t)+\sqrt{\kappa_{2}}c_{in_{2}}(t),
\end{equation}
   
where, $\kappa_{1}$ and $\kappa_{2}$ are the photon decay rates for the optical modes $c_{1}$ and $c_{2}$ respectively. The mechanical oscillator is connected to a thermal bath at a damping rate $\Gamma_{m}$ with a bath occupation (or mean thermal excitation number) given by $n_{th}=\left[exp\left\lbrace \frac{\hbar \omega_{m}}{k_{B}T} \right\rbrace -1 \right]^{-1}$, where $k_{B}$ is the Boltzmann constant and $T$ is the temperature of the mechanical bath. The mechanical mode is also affected by a random Brownian force with zero mean value $\xi_{m}(t)$. Moreover, $c_{in_{1}}(t)$ and $c_{in_{2}}(t)$ represent the input noise operators for the two optical modes, satisfying the following correlation functions \citep{walls}:

\begin{equation}\label{n1}
<c_{in_{1}}(t)c_{in_{1}}(t')>=<c_{in_{1}}^{\dagger}(t)c_{in_{1}}^{\dagger}(t')>=<c_{in_{1}}^{\dagger}(t)c_{in_{1}}(t')>=0,
\end{equation}

\begin{equation}\label{n2}
<c_{in_{2}}(t)c_{in_{2}}(t')>=<c_{in_{2}}^{\dagger}(t)c_{in_{2}}^{\dagger}(t')>=<c_{in_{2}}^{\dagger}(t)c_{in_{2}}(t')>=0,
\end{equation}

\begin{equation}\label{n3}
<c_{in_{1}}(t)c_{in_{1}}^{\dagger}(t')>=<c_{in_{2}}(t)c_{in_{2}}^{\dagger}(t')>=\delta(t-t').
\end{equation}

We are actually interested in analyzing the dynamics of the quantum fluctuations of the system around the steady state in order to establish the presence of quantum correlations among the mechanical mode and the two optical modes at the steady state. To this end, we linearize the quantum Langevin eqns.(\ref{b})-(\ref{c2}) by rewriting each Heisenberg operator as a sum of its steady-state value and a small fluctuation as, $b(t)=\beta_{s}+\delta b(t)$ and $c_{1,2}(t)=\alpha_{1s,2s}+\delta c_{1,2}(t)$. Here, the steady state parameters $\beta_{s}$, $\alpha_{1s}$ and $\alpha_{2s}$ are the solutions of nonlinear algebraic equations obtained by factorizing eqns.(\ref{b})-(\ref{c2}) and setting their time derivatives to zero, given as:

\begin{equation}
\beta_{s}=\frac{i \eta_{1} \alpha_{1s}^{\dagger} \alpha_{1s}+i \eta_{2} \alpha_{2s}^{\dagger} \alpha_{2s}-i \eta_{0}( \alpha_{1s} \alpha_{2s}^{\dagger}+\alpha_{2s} \alpha_{1s}^{\dagger})}{(i\omega_{m}+\Gamma_{m})},
\end{equation}

\begin{equation}
\alpha_{1s}=\frac{2i \eta_{0} \alpha_{2s}Re[\beta_{s}]}{\left[ 2i \eta_{1}Re[\beta_{s}]-(i\omega_{c_{1}}+\frac{\kappa_{1}}{2})\right] },
\end{equation}

\begin{equation}
\alpha_{2s}=\frac{2i \eta_{0} \alpha_{1s}Re[\beta_{s}]}{\left[ 2i \eta_{2}Re[\beta_{s}]-(i\omega_{c_{2}}+\frac{\kappa_{2}}{2})\right] }.
\end{equation}

Thus, by eliminating the steady-state contribution and introducing the amplitude and phase quadratures for the system as $\delta q_{m}(t)=[\delta b(t)+ \delta b^{\dagger}(t)]/\sqrt{2}$, $\delta p_{m}(t)=i[\delta b^{\dagger}(t)- \delta b(t)]/\sqrt{2}$, $\delta X_{c_{1}}(t)=[\delta c_{1}(t)+ \delta c_{1}^{\dagger}(t)]/\sqrt{2}$, $\delta Y_{c_{1}}(t)=i[\delta c_{1}^{\dagger}(t)- \delta c_{1}(t)]/\sqrt{2}$, $\delta X_{c_{2}}(t)=[\delta c_{2}(t)+ \delta c_{2}^{\dagger}(t)]/\sqrt{2}$, $\delta Y_{c_{2}}(t)=i[\delta c_{2}^{\dagger}(t)- \delta c_{2}(t)]/\sqrt{2}$, $X_{in_{1}}(t)=[c_{in_{1}}(t)+c_{in_{1}}^{\dagger}(t)]/\sqrt{2}$, $Y_{in_{1}}(t)=i[c_{in_{1}}^{\dagger}(t)-c_{in_{1}}(t)]/\sqrt{2}$, $X_{in_{2}}(t)=[c_{in_{2}}(t)+c_{in_{2}}^{\dagger}(t)]/\sqrt{2}$ and $Y_{in_{2}}(t)=i[c_{in_{2}}^{\dagger}(t)-c_{in_{2}}(t)]/\sqrt{2}$, we obtain the following linearized equations of motion for the fluctuations of the quadrature operators:

\begin{equation}\label{x1}
\delta \dot{q_{m}}(t)=\omega_{m} \delta p_{m}(t),
\end{equation}

\begin{equation}
\delta \dot{p_{m}}(t)=-\omega_{m} \delta q_{m}(t)-\Gamma_{m} \delta p_{m}(t)+\chi_{1} \delta X_{c_{1}}(t)+\chi_{2} \delta X_{c_{2}}(t)+W_{m}(t),
\end{equation}

\begin{equation}
\delta \dot{X_{c_{1}}}(t)=-\Omega_{c_{1}} \delta Y_{c_{1}}(t)+\eta \delta Y_{c_{2}}(t)-\frac{\kappa_{1}}{2} \delta X_{c_{1}}(t)+\sqrt{\kappa_{1}} X_{in_{1}}(t),
\end{equation}

\begin{equation}
\delta \dot{Y_{c_{1}}}(t)=\Omega_{c_{1}} \delta X_{c_{1}}(t)+ \chi_{1} \delta q_{m}(t)-\eta \delta X_{2}(t)-\frac{\kappa_{1}}{2} \delta Y_{1}(t)+\sqrt{\kappa_{1}} Y_{in_{1}}(t),
\end{equation}

\begin{equation}
\delta \dot{X_{c_{2}}}(t)=-\Omega_{c_{2}} \delta Y_{c_{2}}(t)+\eta \delta Y_{c_{1}}(t)-\frac{\kappa_{2}}{2} \delta X_{c_{2}}(t)+\sqrt{\kappa_{2}}X_{in_{2}}(t),
\end{equation}

\begin{equation}\label{x6}
\delta \dot{Y_{c_{2}}}(t)=\Omega_{c_{2}} \delta X_{c_{2}}(t)+\chi_{2} \delta q_{m}(t)-\eta \delta X_{c_{1}}(t)-\frac{\kappa_{2}}{2} \delta Y_{c_{2}}(t)+\sqrt{\kappa_{2}}Y_{in_{2}}(t),
\end{equation}

with the effective frequencies $\Omega_{c_{1}}=2 \beta_{s} \eta_{1}-\omega_{c_{1}}$ and $\Omega_{c_{2}}=2 \beta_{s} \eta_{2}-\omega_{c_{2}}$ for the cavity modes $c_{1}$ and $c_{2}$ respectively. Here, $\chi_{1}=\eta_{1} \alpha_{1s}-\eta_{0} \alpha_{2s}$ is the effective coupling between the mechanical mode and the optical mode $c_{1}$. However, the effective coupling between the mechanical mode and the optical mode $c_{2}$ is denoted by $\chi_{2}=\eta_{2} \alpha_{2s}-\eta_{0} \alpha_{1s}$. The effective coupling between the two optical modes is given by $\eta=2 \eta_{0} \beta_{s}$. Moreover, $W_{m}(t)=i\sqrt{\frac{\Gamma_{m}}{2}}(\xi_{m}^{\dagger}(t)-\xi_{m}(t))$ is the Hermitian Brownian noise operator, which satisfies the following correlation \citep{gio,gio1}:

\begin{equation}\label{brown}
\left\langle W_{m}(t)W_{m}(t') \right\rangle = \frac{\Gamma_{m}}{\omega_{m}} \int \frac{d\omega}{2\pi} e^{-i\omega(t-t')} \omega\left[1+\coth\left(\frac{\hbar\omega}{2k_{B}T} \right)  \right].
\end{equation}

Brownian noise is the random thermal noise which arises from the stochastic motion of the mechanical oscillator and it is non-Markovian in nature (neither its commutator nor its correlation function is proportional to Dirac delta). This non-Markovian nature of Brownian noise guarantees that the quantum Langevin equations for the system preserve the correct commutation relations between operators during the time evolution \citep{gio1}. The system of above linearized equations of motion (\ref{x1})-(\ref{x6}) can be written in the following compact matrix form \citep{genes}:

\begin{equation}\label{x7}
\dot{R}(t)=M R(t)+N(t),
\end{equation}

where $R(t)=(\delta q_{m}(t), \delta p_{m}(t), \delta X_{c_{1}}(t), \delta Y_{c_{1}}(t), \delta X_{c_{2}}(t), \delta Y_{c_{2}}(t))^{\top}$ (the superscript $\top$ denotes the transposition) is the vector of the quadrature fluctuations. Furthermore, $N(t)=(0,W_{m}(t), \sqrt{\kappa_{1}}X_{in_{1}}(t), \sqrt{\kappa_{1}}Y_{in_{1}}(t), \sqrt{\kappa_{2}}X_{in_{2}}(t), \sqrt{\kappa_{2}}Y_{in_{2}}(t))^{\top}$ is the corresponding vector of noises, whereas $M$ is the drift matrix, given as:

\begin{equation}
         M=
            \left[ {\begin{array}{cccccc}
              0 & \omega_{m} & 0 & 0 & 0 & 0 \\
              -\omega_{m} & -\Gamma_{m} & \chi_{1} & 0 & \chi_{2} & 0 \\
              0 & 0 & -\kappa_{1}/2 & -\Omega_{c_{1}} & 0 & \eta \\
              \chi_{1} & 0 & \Omega_{c_{1}} & -\kappa_{1}/2 & -\eta & 0 \\
              0 & 0 & 0 & \eta & -\kappa_{2}/2 & -\Omega_{c_{2}} \\
             \chi_{2} & 0 & -\eta & 0 & \Omega_{c_{2}} & -\kappa_{2}/2 \\
                \end{array} } \right].
\end{equation} 

The formal solution of eqn.(\ref{x7}) is given by $R(t)=F(t)R(0)+ \int \limits_{0}^{t} ds F(s)N(t-s)$ with $F(t)=e^{Mt}$. The system reaches a steady state only if it is stable, which is possible when all the eigen values of the drift matrix $M$ have negative real parts so that $F(\infty)=0$. In this situation, the stability conditions given in Appendix A must always be satisfied.

Since the quantum noises are white in nature and the dynamics is linearized, hence the steady state of the system will be a zero mean Gaussian state, and therefore can be completely characterized by its $6\times 6$ correlation matrix $V_{ij}=(<R_{i}(\infty)R_{j}(\infty)+R_{j}(\infty)R_{i}(\infty)>)/2$. When the system is stable, starting from the formal solution of eqn.(\ref{x7}), one arrives at

\begin{equation}\label{x8}
V_{ij}(\infty)=\sum_{k,l}\int \limits_{0}^{\infty} ds \int \limits_{0}^{\infty} ds' F_{ik}(s)F_{jl}(s')D_{kl}(s-s'),
\end{equation}   
 
where $D_{kl}(s-s')=(<N_{k}(s)N_{l}(s')+N_{l}(s')N_{k}(s)>)/2$ represents the matrix of the stationary noise correlation functions. Further note that the oscillators with a very high mechanical quality factor $Q=\omega_{m}/\Gamma_{m}\rightarrow \infty$ can only be used in order to achieve the mechanical entanglement. In this weak damping limit $\Gamma_{m} \rightarrow 0$, the quantum Brownian noise becomes $\delta$-correlated \citep{benguria} such that

\begin{equation}
\left\langle W_{m}(t)W_{m}(t')+W_{m}(t')W_{m}(t)\right\rangle \simeq \Gamma_{m}(2n_{th}+1)\delta(t-t'), 
\end{equation}

and one recovers a Markovian process. Within this Markovian approximation of the thermal noise on the mechanical resonator, we finally get $D_{kl}(s-s')=D_{kl}\delta(s-s')$, with the diffusion matrix $D$ given as:

\begin{equation}
D=
    \left[ {\begin{array}{cccccc}
     0 & 0 & 0 & 0 & 0 & 0 \\
     0 & \Gamma_{m}(2n_{th}+1) & 0 & 0 & 0 & 0 \\
     0 & 0 & \kappa_{1}/2 & 0 & 0 & 0 \\
     0 & 0 & 0 & \kappa_{1}/2 & 0 & 0 \\
     0 & 0 & 0 & 0 & \kappa_{2}/2 & 0 \\
     0 & 0 & 0 & 0 & 0 & \kappa_{2}/2 \\
     \end{array} } \right],
\end{equation} 

which is obtained using the definitions of $X_{in_{1}}(t)$, $Y_{in_{1}}(t)$, $X_{in_{2}}(t)$, $Y_{in_{2}}(t)$, eqns.(\ref{n1})-(\ref{n3}) and the fact that the five components of $N(t)$ are uncorrelated. As a consequence, eqn.(\ref{x8}) becomes:

\begin{equation}\label{x9}
V=\int \limits_{0}^{\infty} ds F(s) D F(s)^{\top},
\end{equation}

which is equivalent to the following Lyapunov equation for the correlation matrix in the steady state $[F(\infty)=0]$:

\begin{equation}\label{x10}
MV+VM^{\top}=-D.
\end{equation}

Eqn.(\ref{x10}) is the linear matrix equation and can be straightforwardly solved for $V$. However, the general exact expression is too cumbersome to be reported here. The correlation matrix can provide all the information about the steady state of the system. In the next section, we compute the stationary entanglement between the different modes of the optomechanical system formed by the two cavity modes and a vibrational mode of the mechanical resonator.

\section{Steady-state entanglement}

In this section, we study the steady-state entanglement of the three possible bipartite subsystems, by quantifying it in terms of the logarithmic negativity $E_{N}$ \citep{adesso,vitali} of bimodal Gaussian states. Measurement of the entanglement between any two modes of the system, requires as to compute $E_{N}$, which is obtained by tracing out the third mode (i.e., removing the rows and columns of $V$ which correspond to the third mode). The reduce state is now fully characterized by the $4 \times 4$ matrix $V'$ and still remains Gaussian. In the CV case, the logarithmic negativity $E_{N}$ can be defined as \citep{adesso,vitali}:

\begin{equation}\label{x11}
E_{N}=\max[0,-\ln 2\mu^{-}],
\end{equation} 

where $\mu^{-}\equiv \frac{1}{\sqrt{2}}\sqrt{A-[A^{2}-4 \det(V')]^{1/2}}$. Here, $A=\sum (V')$ with $\sum (V')\equiv \det(X)+\det(Y)-2\det(Z)$, which is the smallest symplectic eigenvalue. We have used the $2 \times 2$ block form of $V'$ as:

\begin{equation}
V'=
    \left[ {\begin{array}{cc}
     X & Z\\
     Z^{\top} & Y\\
     \end{array} } \right].
\end{equation}

Eqn.(\ref{x11}) clearly shows that the logarithmic negativity is the decreasing function of $\mu^{-}$ which basically measures how much two Gaussian states are entangled. A Gaussian state is entangled only if $\mu^{-}<1/2$ (or $4\det(V')<\sum (V')-1/4$), and it is equivalent to Simon's necessary and sufficient entanglement nonpositive partial transpose criterion of the Gaussian states \citep{simon}. However, the second eigenvalue $\mu^{+}\equiv \frac{1}{\sqrt{2}}\sqrt{A-(A^{2}-4\det(A))^{1/2}} \gg 1/2$ at any value of the parameters, thus, it does not affect the nonseparability of the state \citep{simon}. We now analyze the stationary entanglement in the three possible bipartitions of the system using the logarithmic negativity $E_{N}$. Here,  $E_{N}^{(1)}$, $E_{N}^{(2)}$ and $E_{N}^{(3)}$ denote the logarithmic negativities for the mechanical mode-optical mode $c_{1}$, mechanical mode-optical mode $c_{2}$ and optical mode $c_{1}$-optical mode $c_{2}$ entanglements respectively.

\begin{figure}[h]
\hspace{-0.0cm}
\begin{tabular}{cc}
\includegraphics [scale=0.80]{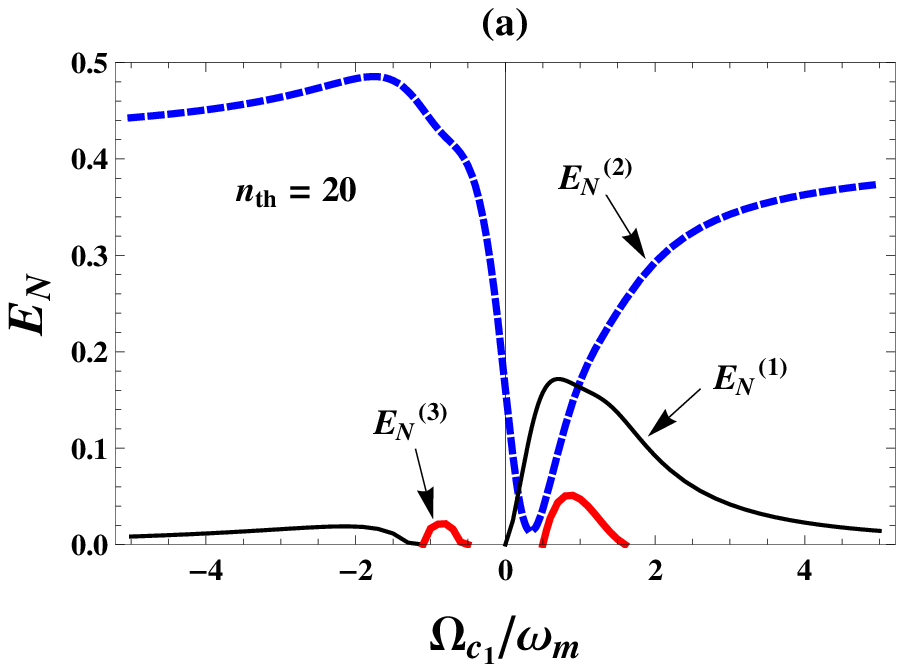}& \includegraphics [scale=0.80]{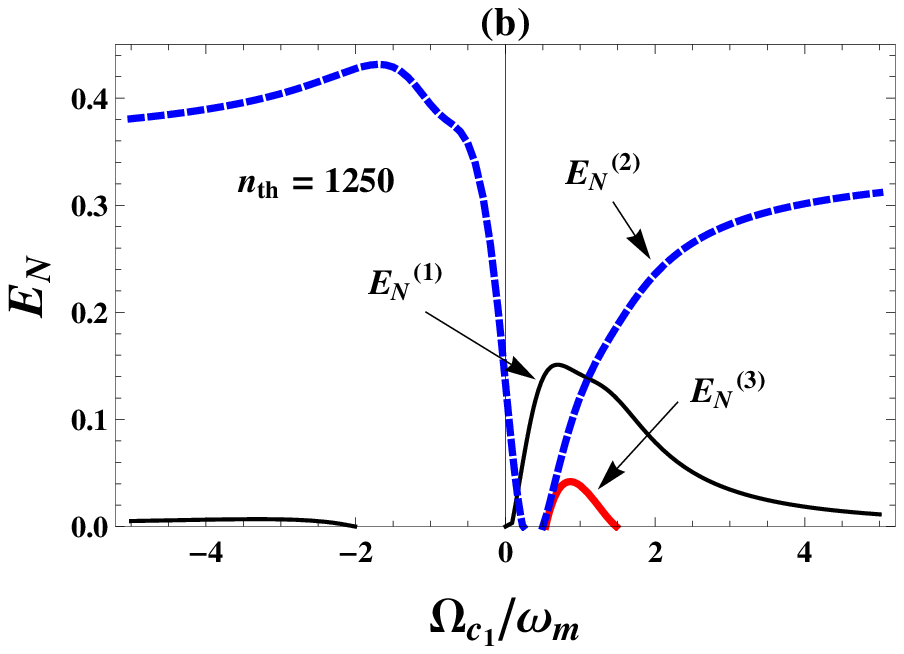}\\
\includegraphics [scale=0.80]{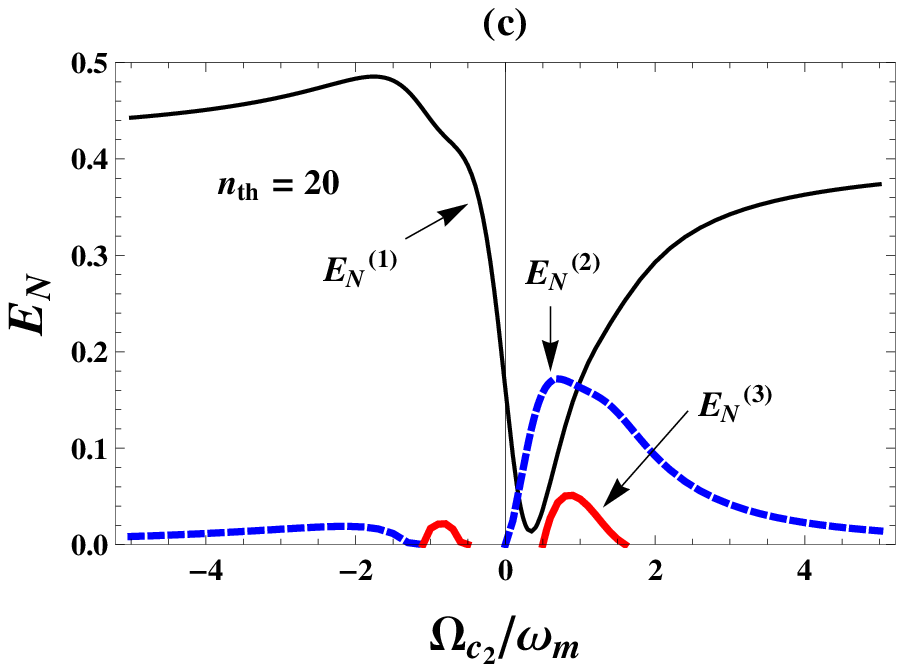}& \includegraphics [scale=0.80]{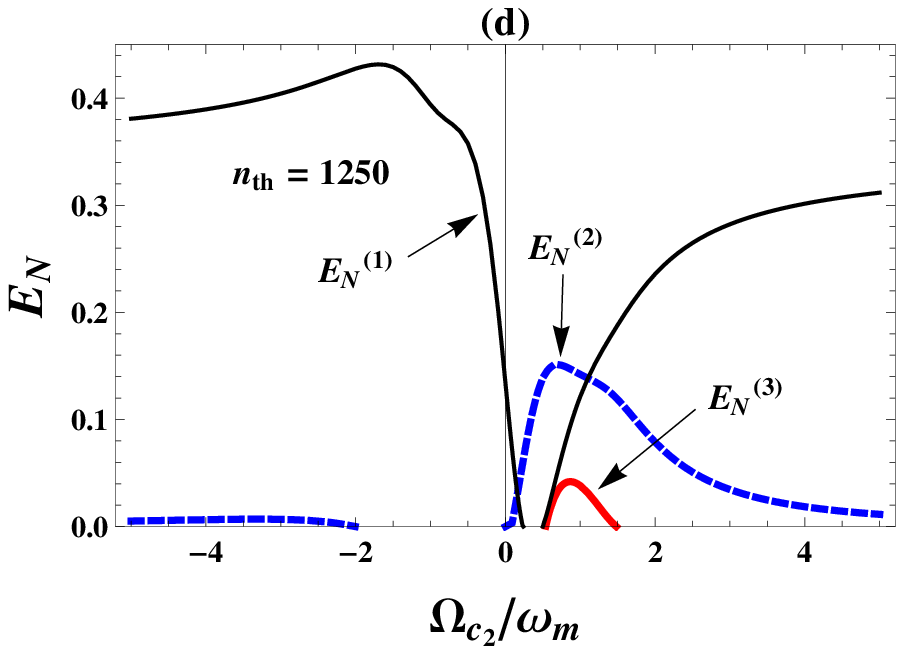}\\
\end{tabular}
\caption{(color online) (a): Plot of logarithmic negativities of the three bipartite cases $E_{N}^{(1)}$ (thin solid line), $E_{N}^{(2)}$ (dashed line) and $E_{N}^{(3)}$ (thick solid line) versus the normalized effective optical frequency ($\Omega_{c_{1}}/\omega_{m}$) for $n_{th}=20$. We have taken $\Gamma_{m}=10^{-5}\omega_{m}$, $\kappa_{1}=\omega_{m}$, $\kappa_{2}=0.5\omega_{m}$, $\chi_{1}=0.1\omega_{m}$, $\chi_{2}=0.9\omega_{m}$, $\eta=0.8\omega_{m}$ and $\Omega_{c_{2}}=-\omega_{m}$. (b): Same as in (a) but for the higher value of bath occupation $n_{th}=1250$. (c): Plot of logarithmic negativities $E_{N}^{(1)}$ (thin solid line), $E_{N}^{(2)}$ (dashed line) and $E_{N}^{(3)}$ (thick solid line) versus the normalized effective optical frequency ($\Omega_{c_{2}}/\omega_{m}$) for $n_{th}=20$. Here, the parameters used are $\kappa_{1}=0.5\omega_{m}$, $\kappa_{2}=\omega_{m}$, $\chi_{1}=0.9\omega_{m}$, $\chi_{2}=0.1\omega_{m}$ and $\Omega_{c_{1}}=-\omega_{m}$. Other parameters are same as in (a). (d): Same as in (c) but for $n_{th}=1250$.}
\end{figure}\label{fig2}

\begin{figure}[h]
\hspace{-0.0cm}
\begin{tabular}{cc}
\includegraphics [scale=0.80]{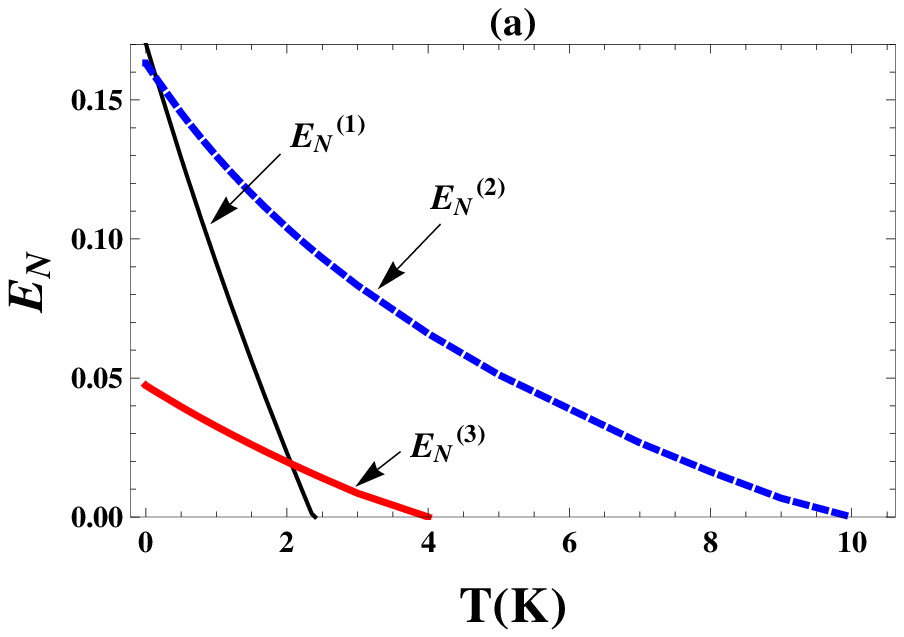}& \includegraphics [scale=0.80]{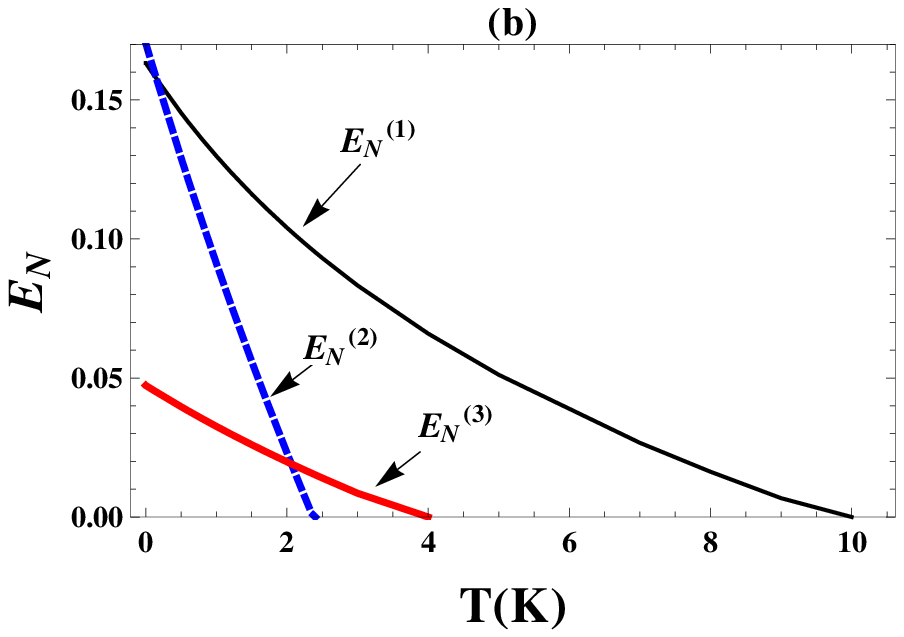}\\
\end{tabular}
\caption{(color online) (a): Plot of logarithmic negativities of the three bipartite cases $E_{N}^{(1)}$ (thin solid line), $E_{N}^{(2)}$ (dashed line) and $E_{N}^{(3)}$ (thick solid line) as a function of bath temperature $T$. The parameter values are $\omega_{m}=2 \pi \times 10^{7}$Hz, $\Gamma_{m}=10^{-5}\omega_{m}$, $\kappa_{1}=0.5\omega_{m}$, $\kappa_{2}=\omega_{m}$, $\chi_{1}=0.9\omega_{m}$, $\chi_{2}=0.1\omega_{m}$, $\eta=0.8\omega_{m}$, $\Omega_{c_{1}}=-\omega_{m}$ and $\Omega_{c_{2}}=\omega_{m}$. (b): Plot of logarithmic negativities $E_{N}^{(1)}$ (thin solid line), $E_{N}^{(2)}$ (dashed line) and $E_{N}^{(3)}$ (thick solid line) as a function of $T$ for $\kappa_{1}=\omega_{m}$, $\kappa_{2}=0.5\omega_{m}$, $\chi_{1}=0.1\omega_{m}$, $\chi_{2}=0.9\omega_{m}$, $\Omega_{c_{1}}=\omega_{m}$ and $\Omega_{c_{2}}=-\omega_{m}$. Other parameters used are same as in (a).}
\end{figure}\label{fig3}

\begin{figure}[h]
\hspace{-0.0cm}
\begin{tabular}{cc}
\includegraphics [scale=0.80]{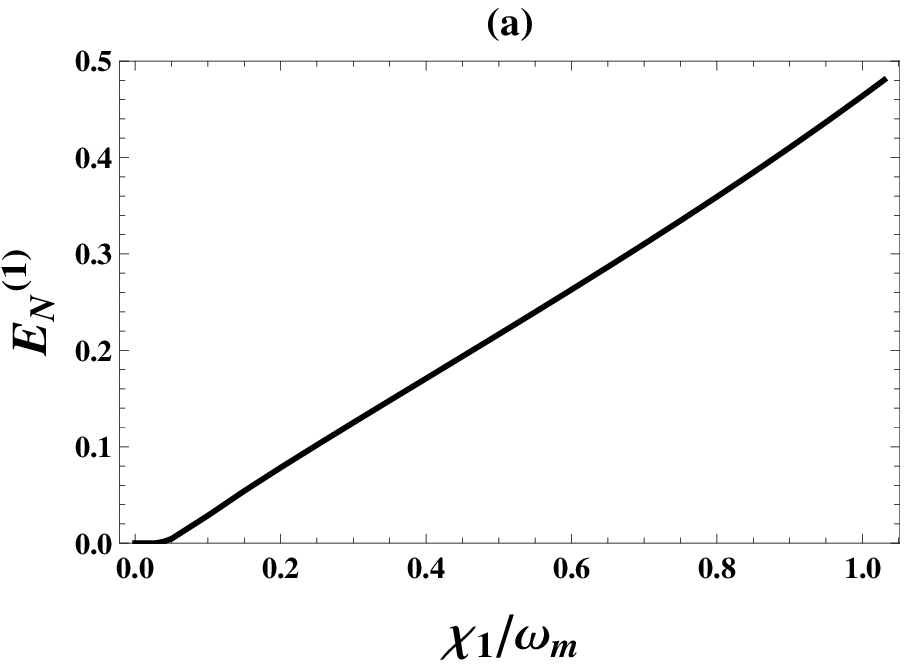}& \includegraphics [scale=0.80]{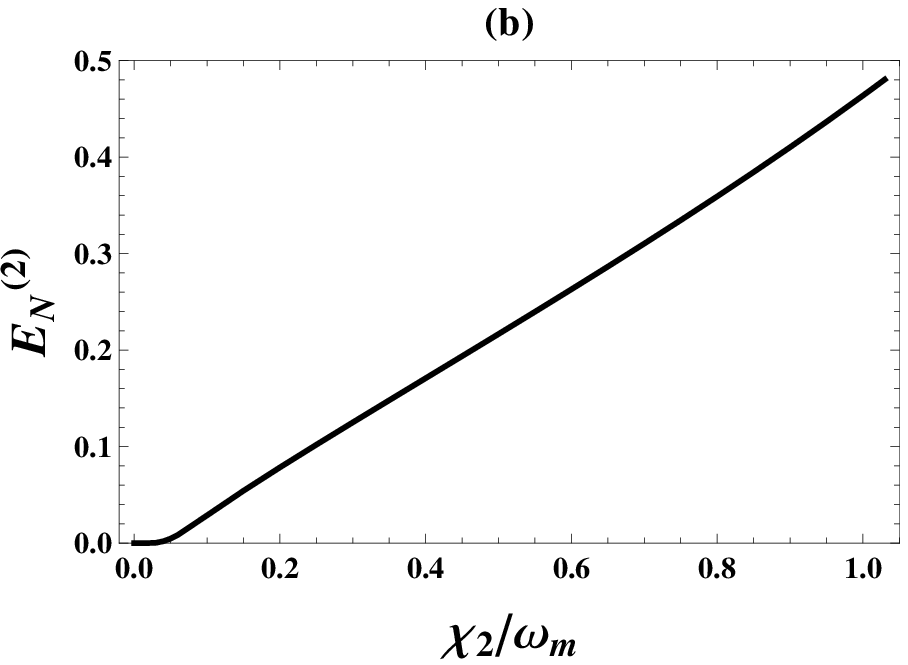}\\
\end{tabular}
\caption{(color online) (a): Plot of logarithmic negativity $E_{N}^{(1)}$ as a function of dimensionless effective coupling parameter $\chi_{1}/\omega_{m}$ for $\chi_{2}=0.01\omega_{m}$ and $\eta=0.01\omega_{m}$. (b): Plot of logarithmic negativity $E_{N}^{(2)}$ as a function of $\chi_{2}/\omega_{m}$ for $\chi_{1}=0.01\omega_{m}$ and $\eta=0.01\omega_{m}$. The other parameters used are $\Gamma_{m}=10^{-5}\omega_{m}$, $\kappa_{1}=0.5\omega_{m}$, $\kappa_{2}=0.5\omega_{m}$, $\Omega_{c_{1}}=-\omega_{m}$, $\Omega_{c_{2}}=-\omega_{m}$ and $n_{th}=20$.}
\end{figure}\label{fig4}

In fig.2(a), the logarithmic negativities of the three bipartite cases $E_{N}^{(1)}$ (solid line), $E_{N}^{(2)}$ (dashed line) and $E_{N}^{(3)}$ (dot dashed line) are plotted as a function of normalized effective optical frequency $\Omega_{c_{1}}/\omega_{m}$ for mean thermal excitation number $n_{th}=20$ (corresponding to a bath temperature $T=0.01 $K). Fig.2(b) shows the same plot but at a higher value of bath occupation, $n_{th}=1250$ (i.e., $T=0.6 $K). We have also shown the logarithmic negativities for the three bipartite entanglements versus normalized effective cavity frequency $\Omega_{c_{2}}/\omega_{m}$ for $n_{th}=20$ (fig.2(c)) and $n_{th}=1250$ (fig.2(d)). The stationary entanglement between a single driven optical cavity field mode and a mechanical resonator via radiation pressure has already been analyzed in \citep{vitali1}. Our model Hamiltonian involves the two cavity modes with different frequencies, each driven by an intense laser, which are not only separately coupled via pondermotive interaction to the mechanical resonator but are also coupled together via mechanical oscillator. Due to the presence of this second optical cavity mode, we find that the steady-state entanglement is now generated within the three subsystems, namely, mechanical mode-optical mode $c_{1}$, mechanical mode-optical mode $c_{2}$ and optical mode $c_{1}$-optical mode $c_{2}$. The simultaneous presence of all the three possible bipartite entanglements in the chosen parameter regime for a wide range of effective optical frequencies ($-5\omega_{m} < \Omega_{c_{1,2}} < 5\omega_{m}$) witnesses the strong correlation between the two optical fields and the mechanical oscillator at the steady state. However, these entanglements vanish at some values of effective optical frequencies. Thus, nonzero entanglements can be obtained by the proper choice of the system parameters. Further note that the presence of nonzero entanglement between the two optical modes is due to the effect of the mechanical motion of the resonator since both the cavity modes are coupled together via mechanical oscillator only. There would be no such entanglement possible if the mechanical element is fixed. Moreover, at a higher temperature $T=0.6$K, the qualitative behaviour of all the three possible bipartite entanglements shown in figs.2(b) and 2(d) is identical to that of the corresponding figs.2(a) and 2(c) respectively. Although, the achievable values of these stationary entanglements are comparatively lower. Despite the lower values, they are still quite robust against temperature. This is studied in more detail in fig.3, where we study the robustness of the steady-state entanglements with respect to reservoir temperature $T$. As expected, the logarithmic negativities $E_{N}^{(1)}$, $E_{N}^{(2)}$ and $E_{N}^{(3)}$, hence the entanglements decay with the increase in mechanical resonator's environmental temperature. Figs.3(a) and 3(b) further depict that the entanglement between the mechanical mode and any of the two optical modes can be selectively made large and more robust against thermal noise by choosing an appropriate parameter regime. Thus, one can not only detect the optomechanical entanglement but can also selectively optimize and increase it. Moreover, for $T<2.4$K, the simultaneous presence of all the three possible bipartite stationary entanglements again confirms the strong correlation between the three bosonic modes.

We further show the variation in optomechanical entanglements with respect to their effective optomechanical couplings in fig.4. Fig.4(a) illustrates the plot of logarithmic negativity $E_{N}^{(1)}$ as a function of dimensionless effective coupling $\chi_{1}/\omega_{m}$. In fig.4(b), logarithmic negativity $E_{N}^{(2)}$ is plotted versus dimensionless effective coupling parameter $\chi_{2}/\omega_{m}$. It can be clearly seen from fig.4(a) that the optomechanical entanglement $E_{N}^{(1)}$ increases monotonically with the increase in coupling. However, rest of the two entanglements $E_{N}^{(2)}$ and $E_{N}^{(3)}$ are always zero in the range of parameters considered here. Similarly, fig.4(b) shows a monotonic increase in $E_{N}^{(2)}$ with increase in $\chi_{2}/\omega_{m}$, such that $E_{N}^{(1)}$ and $E_{N}^{(3)}$ remain zero for all the values of $\chi_{2}/\omega_{m}$. However, the stability conditions always put some upper bound on the maximum achievable value of optomechanical coupling, which can be improved without entering the unstable regime by using the high cavity finesse. Further note that the entanglement between two optical modes is always zero for all the values of effective optical coupling $\eta$ with $\chi_{1}=\chi_{2}=0.01\omega_{m}$ by keeping rest of the parameters same as in fig.4. This is due to the fact that the two optical modes are indirectly coupled via mechanical resonator and can only form a Gaussian entangled state for sufficiently high values of optomechanical couplings as shown in fig.2.  

In order to demonstrate that the dynamics investigated here are within the experimental reach, we now discuss the experimental prospects for various parameters used in the main paper. In our calculations, the frequency of mechanical oscillator is taken to be $2 \pi \times 10^{7}$Hz with a mechanical quality factor of $10^{5}$, which is very close to that of recently performed experiments \citep{anetsberger,arcizet,gigan}. Advances in design, fabrication and material properties are expected to lead to high-finesse optical cavities with a decay rate of nearly $2 \pi \times 10$MHz \citep{tanaka,notomi}. The coupling rate in optomechanical crystal setups is currently recorded to be $2 \pi \times 1$MHz \citep{chan}. Utilizing nanoslots \citep{robinson} to enhance the local optical field in such structures can lead to the coupling rates above $10$MHz. In typical optomechanical experiments, the limit $\hbar \Gamma_{m} << \hbar \omega_{m} << k_{B} T$ is always taken into account \citep{hadjar,tittonen,cohadon,pinard}.

Now, we discuss the optomechanical quantum nondemolition (QND) phonon and photon detection scheme. For the QND measurement of phonon and photon number, the two optical cavity modes ($c_{1}$ and $c_{2}$) should be assumed to be independently driven by using the two separate laser sources having frequencies $\omega_{l_{1}}$ and $\omega_{l_{2}}$ respectively. The transmitted signal from each of the cavity modes should be assumed to be filtered and measured independently using two separate photodetectors $D_{1}$ and $D_{2}$ respectively \citep{ludwig}. In principle, one could utilize this spectral filtering to perform the selective mode coupling \citep{safavi}. As discussed in \citep{ludwig}, the photon number of the cavity mode $c_{1}$ can be detected by using the another cavity mode $c_{2}$. The two independently driven optical cavity modes are chosen in order to suppress the influence of unwanted transitions from the detection mode ($c_{2}$) to the signal mode ($c_{1}$). The information about the photon number of the cavity mode $c_{1}$ can be extracted by using the data from the photodetector $D_{2}$. Here, it is assumed that the detection mode has a lower finesse than the signal mode such that a sufficiently large number of photons arrives at the photodetector $D_{2}$, whereas, the state of $c_{1}$ is only weakly perturbed by the photons in $c_{2}$. Moreover, in order to detect the photon number within its lifetime, it is also required that the measurement time should always be less than the inverse of the product of cavity decay rate and the photon number. During the measurement time, there should be no excitation of phonon. For the optomechanical QND phonon detection, one of the cavity modes ($c_{1}$) is pumped with a laser at frequency $\omega_{l_{1}}$ and the transmitted signal is measured using a photodetector $D_{1}$. The second cavity mode $c_{2}$ remains undriven and behaves like an idle spectator. Here, the photon number in the detection mode $c_{1}$ corresponding to different phonon states is to be studied firstly as described in \citep{ludwig}. This detection mode photon number should follow the time evolution of the mechanical mode. In this way, the continuous monitoring of photon counts at the photodetector gives the QND measurement of the phonon number. The phonon detection using the second optical cavity mode $c_{2}$ can be described analogously.

\section{Conclusion}

In conclusion, we have shown that an optomechanical system formed by two optical cavity modes and a mechanical mode can produce stationary three possible bipartite entanglements in an experimentally accessible parameter regime. In this regime, it was also observed that the mechanical mode and the two optical modes of the system are strongly correlated for the temperatures below $T=2.4$K. Such a strongly coupled three-mode system showing robust steady-state entanglements against temperature can be exploited for the realization of quantum memories and quantum interfaces within quantum-communication networks. We have also observed that the entanglement between the mechanical mode and any of the two optical modes of the system can be selectively made large and more robust against temperature by the proper choice of the system parameters. It was further seen that the optomechanical entanglement increases with the increase in optomechanical coupling such that the optimal entanglement is achieved for the largest coupling allowed by the stability condition. Moreover, despite of appreciably high optical coupling, we found that the nonzero entanglement between the two optical modes is only possible for sufficiently high values of optomechanical couplings since both the cavity modes are indirectly coupled via mechanical oscillator. 

\section{Acknowledgements}

Neha Aggarwal and A. Bhattacherjee acknowledge financial support from the Department of Science and Technology, New Delhi for financial assistance vide grant SR/S2/LOP-0034/2010. Sonam Mahajan acknowledges University of Delhi for the University Teaching Assistantship.

\section{Appendix A}

The stability conditions for the system can be obtained by applying the Routh-Hurwitz criterion \citep{gradshteyn,dejesus}, which gives the following two nontrivial stability expressions on the system parameters:

\begin{equation}
S_{1}=a_{0}>0,
\end{equation}

\begin{equation}
S_{2}=(a_{5}a_{4}a_{3}+a_{6}a_{1}a_{5}-a_{6}a_{3}^{2}-a_{2}a_{5}^{2})>0,
\end{equation}

where

\begin{eqnarray}
\begin{split}
a_{0}&=\frac{\kappa_{1}^{2}\kappa_{2}^{2}\omega_{m}^{2}}{16}+\frac{\eta^{2}\omega_{m}^{2}\kappa_{1}\kappa_{2}}{2}-\frac{\omega_{m}\eta \chi_{1}\chi_{2}\kappa_{1}\kappa_{2}}{2}+\frac{\kappa_{1}^{2}\omega_{m}^{2}\Omega_{c_{2}}^{2}}{4}+\frac{\kappa_{1}^{2}\omega_{m}\Omega_{c_{2}}\chi_{2}^{2}}{4}+\frac{\kappa_{2}^{2}\omega_{m}^{2}\Omega_{c_{1}}^{2}}{4}\\ &+\frac{\kappa_{2}^{2}\omega_{m}\Omega_{c_{1}}\chi_{1}^{2}}{4}+\omega_{m}^{2}(\Omega_{c_{1}}^{2}\Omega_{c_{2}}^{2}+\eta^{4}-2\eta^{2}\Omega_{c_{1}}\Omega_{c_{2}})
+\omega_{m}(\chi_{1}^{2}\Omega_{c_{2}}^{2}\Omega_{c_{1}}-\chi_{1}^{2}\eta^{2}\Omega_{c_{2}})
\\ &+\omega_{m}(2\eta \chi_{1}\chi_{2}\Omega_{c_{1}}\Omega_{c_{2}}
-2\chi_{1}\chi_{2}\eta^{3}+\chi_{2}^{2}\Omega_{c_{1}}^{2}\Omega_{c_{2}}
-\Omega_{c_{1}}\chi_{2}^{2}\eta^{2}),
\end{split}
\end{eqnarray}

\begin{eqnarray}
\begin{split}
a_{1}&=\Gamma_{m}\left[\frac{\kappa_{1}^{2}\kappa_{2}^{2}}{16}+\frac{\eta^{2}\kappa_{1}\kappa_{2}}{2}+\frac{\kappa_{1}^{2}\Omega_{c_{2}}^{2}}{4}+\frac{\kappa_{2}^{2}\Omega_{c_{1}}^{2}}{4}+\Omega_{c_{1}}^{2}\Omega_{c_{2}}^{2}+\eta^{4}
-2\eta^{2}\Omega_{c_{1}}\Omega_{c_{2}}\right]\\ &+\frac{\omega_{m}^{2}\kappa_{1}\kappa_{2}}{4}(\kappa_{1}+\kappa_{2})+(\kappa_{1}+\kappa_{2})(\eta^{2}\omega_{m}^{2}-\omega_{m}\eta\chi_{1}\chi_{2})\\ &+\kappa_{1}(\omega_{m}^{2}\Omega_{c_{2}}^{2}+\omega_{m}\Omega_{c_{2}}\chi_{2}^{2})+\kappa_{2}(\omega_{m}^{2}\Omega_{c_{1}}^{2}+\omega_{m}\Omega_{c_{1}}\chi_{1}^{2}), 
\end{split}
\end{eqnarray}

\begin{eqnarray}
\begin{split}
a_{2}&=\frac{\kappa_{1}^{2}\kappa_{2}^{2}}{16}+\omega_{m}^{2}\left[\frac{\kappa_{1}^{2}}{4}+\frac{\kappa_{2}^{2}}{4}+\kappa_{1}\kappa_{2} \right]+\frac{\kappa_{1}^{2}\kappa_{2}\Gamma_{m}}{4}+\frac{\kappa_{2}^{2}\kappa_{1}\Gamma_{m}}{4}+2\eta^{2}\omega_{m}^{2}-2\omega_{m}\eta \chi_{1}\chi_{2}\\ &+\eta^{2}\left[\frac{\kappa_{1}\kappa_{2}}{2}+\Gamma_{m}\kappa_{1}+\Gamma_{m}\kappa_{2} \right]+\Omega_{c_{2}}^{2}\left[\frac{\kappa_{1}^{2}}{4}+\omega_{m}^{2}+\kappa_{1}\Gamma_{m} \right]+\eta^{4}-2\eta^{2}\Omega_{c_{1}}\Omega_{c_{2}}\\ &+\Omega_{c_{1}}^{2}\left[\frac{\kappa_{2}^{2}}{4}+\omega_{m}^{2}+\kappa_{2}\Gamma_{m} \right]+\omega_{m}\Omega_{c_{2}}\chi_{2}^{2}+\omega_{m}\Omega_{c_{1}}\chi_{1}^{2}
+\Omega_{c_{1}}^{2}\Omega_{c_{2}}^{2},  
\end{split}
\end{eqnarray}

\begin{eqnarray}
a_{3}=\kappa_{1}\left[\frac{\kappa_{2}^{2}}{4}+\omega_{m}^{2}+\eta^{2}+\Omega_{c_{2}}^{2} \right]+\kappa_{2}\left[\frac{\kappa_{1}^{2}}{4}+\omega_{m}^{2}+\eta^{2}+\Omega_{c_{1}}^{2} \right]+\Gamma_{m}\left[\frac{\kappa_{1}^{2}}{4}+\frac{\kappa_{2}^{2}}{4}+\kappa_{1}\kappa_{2}+2\eta^{2}+\Omega_{c_{1}}^{2}+\Omega_{c_{2}}^{2} \right],  
\end{eqnarray}

\begin{eqnarray}
a_{4}=\frac{\kappa_{1}^{2}}{4}+\frac{\kappa_{2}^{2}}{4}+\kappa_{1}\kappa_{2}+\Gamma_{m}(\kappa_{1}+\kappa_{2})+\omega_{m}^{2}+\Omega_{c_{1}}^{2}+\Omega_{c_{2}}^{2}+2\eta^{2},
\end{eqnarray}

\begin{eqnarray}
a_{5}=(\Gamma_{m}+\kappa_{1}+\kappa_{2}),
\end{eqnarray}

\begin{equation}
a_{6}=1.
\end{equation}

\end{document}